\documentclass[12pt]{article}
\usepackage{amssymb,amsmath,amsfonts,multirow,rotate,color,graphicx}
\usepackage{hyperref}

\usepackage{tikz,lipsum,lmodern}
\usepackage[most]{tcolorbox}

\usepackage{graphicx}

\usepackage{times}

\usepackage{booktabs}

\newenvironment{sciabstract}{%
\begin{quote} \bf}
{\end{quote}}

\newcounter{lastnote}

\usepackage{xcolor}
\definecolor{RoyalBlue}{HTML}{4169e1}
\definecolor{ForestGreen}{HTML}{228b22}

\usepackage{setspace}
\usepackage{subfigure}
\usepackage{comment}

\usepackage{todonotes}

\newcommand{\rev}[1]{{\color{black} #1}}

\title{The resilience of the multirelational structure of geopolitical treaties is critically linked to past colonial world order and offshore fiscal havens}

\author
{Pier Luigi Sacco$^{1,2\ast}$, Alex Arenas$^{3}$, Manlio De Domenico $^{4\ast}$\\
\\
\normalsize{$^{1}$DiSFiPEQ, University of Chieti-Pescara,}
\normalsize{Viale Pindaro 42, 65127 Pescara, Italy}\\
\normalsize{$^{2}$metaLAB (at) Harvard,}
\normalsize{42 Kirkland St, 20138 Cambridge MA}\\
\normalsize{$^{3}$School of Computer Science and Mathematics, Universitat Rovira i Virgili,} \\
\normalsize{Av.da Paisos Catalans 26, 43004 Tarragona (Spain)}\\
\normalsize{$^{4}$Department of Physics and Astronomy ``Galileo Galilei'', University of Padova, Padova, Italy}\\ 
\normalsize{Via Francesco Marzolo 8, 35121 Padova, Italy}
\\
\footnotesize{$^\ast$Corresponding author. E-mail:  manlio.dedomenico@unipd.it; pierluigi\_sacco@fas.harvard.edu}
}

\date{}

\begin{document} 

\baselineskip24pt

\maketitle


\begin{sciabstract} 
The governance of the political and economic world order builds on a complex architecture of international treaties at various geographical scales. In a historical phase of high institutional turbulence, assessing the stability of such architecture with respect to the unilateral defection of single countries and to the breakdown of single treaties is important. We carry out this analysis on the whole global architecture and find that the countries with the highest disruption potential are not major world powers (with the exception of Germany and France) but mostly medium-small and micro-countries. Political stability is highly dependent on many former colonial overseas territories that are today part of the global network of fiscal havens, as well as on emerging economies, mostly from South-East Asia. Economic stability depends on medium sized European and African countries. However, single global treaties have surprisingly less disruptive potential, with the major exception of the WTO. These apparently counter-intuitive results highlight the importance to a nonlinear approach to international relations where the complex multilayered architecture of global governance is analyzed using advanced network science techniques. Our results suggest that the potential fragility of the world order seem to be much more directly related to global inequality and fiscal injustice than it is commonly believed, and that the legacy of the colonial world order is still very strong in the current international relations scenario. In particular, vested interests related to tax avoidance seem to have a structural role in the political architecture of global governance.
\end{sciabstract}




The global architecture of flows of people, goods,  services and information is regulated by a governance system that works through a complex web of institutions and international treaties, most of which have been developed in the postwar period to promote an enabling \rev{environment} \cite{mcgrew2002governing} to exploit opportunities of various nature, such as, for instance, gains from trade, optimal division of labor, free circulation of people and resources, reduction of uncertainty from local economic fluctuations, multi-lateral conflict prevention and resolution arrangements, and so on. Taken together, such agreements form a complex, multi-layered system, often associated to the notion of a 'world order' \cite{kissinger2014world}.

The system has been mostly tailored to the challenges and needs of a bipolar, cold war world which called for an overarching institutional setting to prevent the escalation of conflict and to mitigate frictional obstacles to trade, movement, cultural exchange, and so on \cite{ashley2007alliance}. The system further developed in the apparently unipolar moment that followed the collapse of the former Soviet Union, but has been increasingly under pressure even since \cite{freeman2021american}, for the concurrence of several critical factors. The first has been the rapid emergence of a new multipolar world order \cite{mead2014return} in which several powers are competing for global influence \cite{duncombe2018after}, including at least USA, the EU, China, India, and Russia \cite{parizek2017representation}, and paving the way to alternative conceptions of the world order itself \cite{agnew2010emerging}, and to new variable geometries of influence \cite{bisley2019contested},  \cite{kaczmarski2015russia}. The second is the change of direction in US foreign policy which has gradually reduced the country's pivotal role in the political and financial maintenance of the global system to increasingly prioritize the internal affairs agenda \cite{patrick2017trump}. The third are major threats such as the climate change emergency and the global pandemic crises \cite{roberts2011multipolarity}, \cite{aziz2021covid}, and previously the great credit crash of 2007-9 which severely tested the resilience of the multilateral architecture of the global order \cite{samman2016conjuring}. In a multipolar order, there is not necessarily an interest in guaranteeing the multilateral functioning of an institutionalized global governance system \cite{litman2017g20}, and some powers could have an interest in disrupting it at least in part to gain more influences in strategic regions, including the US themselves in the light of the new political agenda. And moreover, the global governance principles that could work in a 20th century setting need not be as effective in a 21st century one \cite{dieter2016g20}, as proven by its basic inability to provide timely and effective responses to the new challenges \cite{cameron2020eu}, whose scale and complexity is unprecedented and calls fore radically new solutions and extreme levels of global institutional coordination and cooperation \cite{conversi2020ultimate}.

In this new context, the possibility of a sudden collapse of some of the institutions and treaties that shape the governance system is no longer a mere theoretical speculation \cite{guven2017world}, and likewise it is possible that even major global powers contemplate their partial or total withdrawal from certain institutions or treaties \cite{da2013clash}, as it has happened for the US under the Trump administration, with the withdrawal from UNESCO and UNHRC as a form of political retaliation \cite{suleman2019trump}, and the threat to leave the WHO, which was not ratified by the incoming Biden administration. Even the historical alliance between the US and major European countries, which is at the root of the very definition of the West as the geo-political cornerstone of the whole system \cite{smith2018eu}, cannot be taken for granted anymore in the new scenario \cite{ikenberry2017plot}, \cite{riddervold2018unified}. Populist governments all over the world build their political agenda on extreme parochialism and opportunistic, case-by-case adherence to international agreements, often as a revenge to perceived political or economic marginality \cite{rodriguez2018revenge}. On the other hand, the pandemic crisis de facto temporarily suspended certain treaties, such as the free intra-EU mobility ensured by the Schengen Treaty during anti-Covid-19 lockdowns, as well as many other free circulation agreements, and such temporary suspensions could pave the way to larger and more stable institutional disruptions if normal functioning cannot be restored quickly.

It is important therefore to analyze the global governance system's robustness against this kind of shocks. Not all institutions and treaties have the same structural role in the global architecture, and the same holds for the participation of a given nation to one or more institutions and treaties. Which are the truly critical ones? To the current state of knowledge, there is no clear answer to this question, despite its undeniable importance. In this paper, we provide a first, systematic analysis of the global governance system against one-sided disruptions, both in terms of treaty suspensions and of countries' unilateral withdrawals. This allows us to derive new risk indicators that may become very relevant in strategic and scenario analysis in the coming years. 

In this increasingly uncertain scenario, interpreting global changes in terms of linear processes of structural change is likely misleading. The tendency to interpret the evolution of the world order as a mere interplay between the goals and interests of the most influential countries and country blocks remains strong. However, a nonlinear science of networked international relations \cite{jervis1998system} is much more appropriate to deal with the complexity of international relations \cite{alter2018rise}, and especially so in response to major, unexpected shocks. We can conceptualize the world order as a multilevel network of alliances between countries, whose structural characteristics critically impinge upon its resilience. Inspired in biology, here we propose a holistic approach to assess the resilience of world order to the breakdown of geopolitical treaties due to unilateral moves by specific players. We call this approach systems foreign policy, in analogy to systems biology \cite{kitano2002computational}, defined as the computational and mathematical modeling of complex geopolitical systems. Given the existing complex architecture of global relations, even minor changes in existing economic or political trade deals could in principle spark complex dynamic responses. One cannot role out in principle that the collapse of relatively minor agreements or the withdrawal of second-tier countries from major agreements kick off adjustment cascades whose consequences could be disruptive. We propose an innovative methodology based upon the structure of multilayer networks \cite{de2013mathematical} \cite{kivela2014multilayer} as a substrate for a nonlinear approach to the analysis of international relations, and we illustrate its potential by simulating the structural impact of simple shocks under the form of unilateral defections on the current world order architecture.

\section*{Results}

\begin{figure}[!t]
\centering
\includegraphics[width=0.5\linewidth]{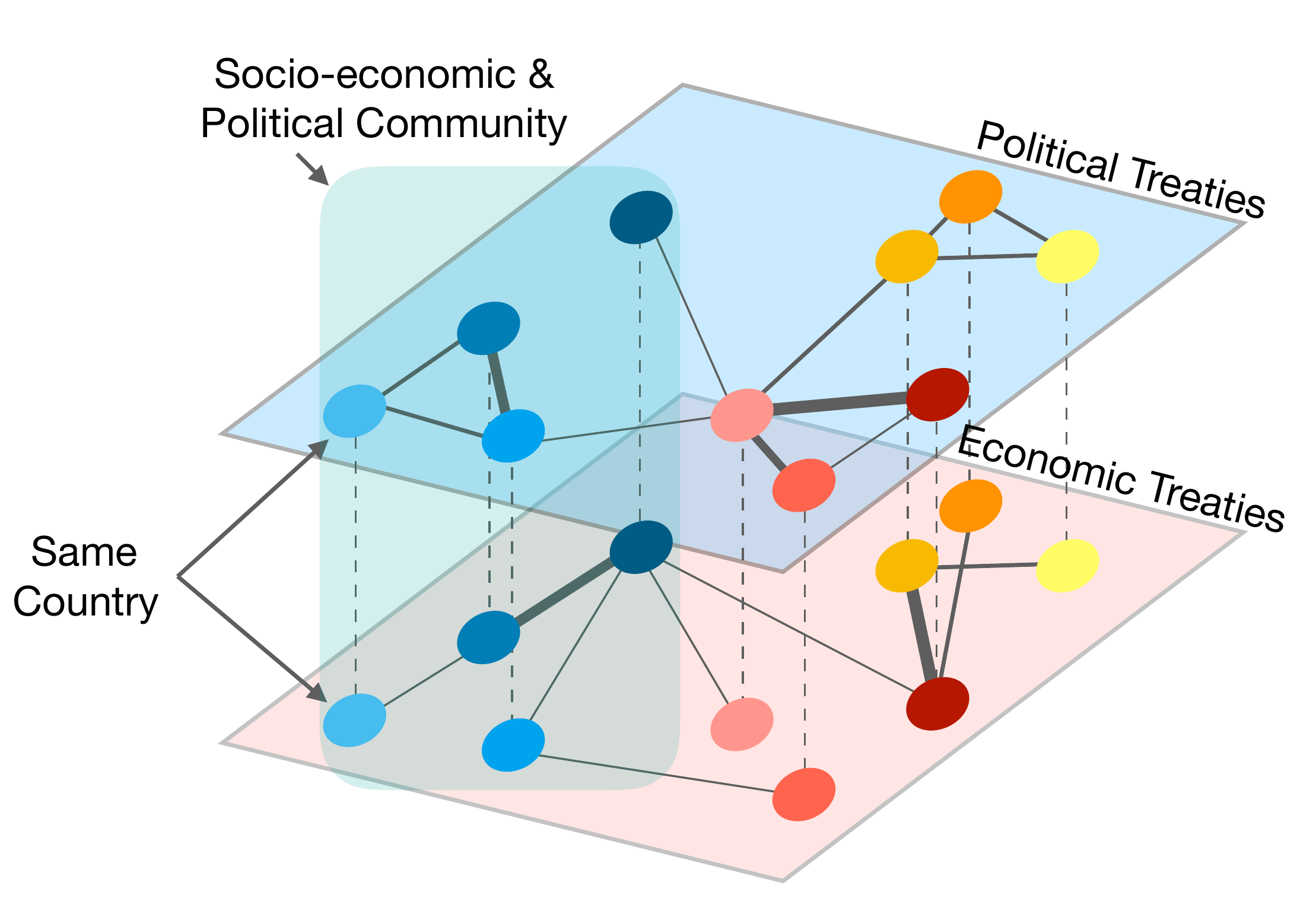}
\caption{\textbf{Multilayer network representation of socio-economic and political treaties.} Nodes represent countries, edges represent existing interactions: different relationships, like political and economic ones, are mapped to different layers.}
\label{fig:diagram}
\end{figure}

\paragraph{Overview of the data.} We consider the full list of active (as of 2015) economic alliances, agreements and bilateral trades, and political alliances and agreements (see Tab.~\ref{tab:acronym}). The data set consists of 200 countries, with 4,733 deals in the political layer, and 14,890 in the economic layer. Only 313 political interactions are not reflected in the economic layer. The resulting multilayer network is abstracted as two layers, political and economic, whose nodes represent countries, and links represents deals in their respective layers (Fig.~\ref{fig:diagram}). 

\paragraph{The damage index.} We test the world order resilience by evaluating the global structural consequences of a unilateral defection of one or more countries from their (economic or political) deals. To this purpose, we first compute the community structure of the multilayer network \cite{de2015identifying}. Communities are detected as groups of countries whose deals are denser between them than with the rest of countries outside their group. Several publicly available algorithms allow to carry out such computation \cite{fortunato2010community}. Next, we simulate the disruption by eliminating a country and all its deals from one of the layers, and re-compute the resulting mesoscale organization, including community structure and seeking for eventual components disconnected from the system's core. From this information, we can define a Damage Index measuring the level of disintegration of the original communities, the emergence of new connected components and the loss of nodes in the largest connected component. 

The mathematical definition of the damage index is easy and intuitive. Let  $C^{\ell}$ be the original number of communities before any disruption, and $C_{i}^{\ell}$ indicate the number of communities found after removing country (or block) $i$ from layer $\ell$. We define the ratio $c_{i}^{\ell} = C_{i}^{\ell}/C^{\ell}$. Similarly, we define the indicators $q_{i}^{\ell}$ and $g_{i}^{\ell}$ as the number of connected components in the system, and the size of the largest connected component, respectively. The three indices, separately, provide complementary information about how a country (or block) removal alters the whole structure, the size of its core and the number of disconnected clusters (note that there can be just one connected component but several communities). The damage index is defined by 
\begin{eqnarray}
\delta_{i}^{\ell} = \frac{c_{i}^{\ell} \times q_{i}^{\ell}}{g_{i}^{\ell}},
\end{eqnarray}
and its normalized version as  $\tilde{\delta}_{i}^{\ell}=\delta_{i}^{\ell}/\max(\delta_{i}^{\ell})$. Note that the damage index is high when, after disruption, the number of communities increases (more segregated network) or/and the number of connected components increases (segregation and isolation), or when the largest component of the system decreases (disaggregation). Note that these variables are not independent but truly correlated.

\begin{figure*}[!t]
\centering
\includegraphics[width=\linewidth]{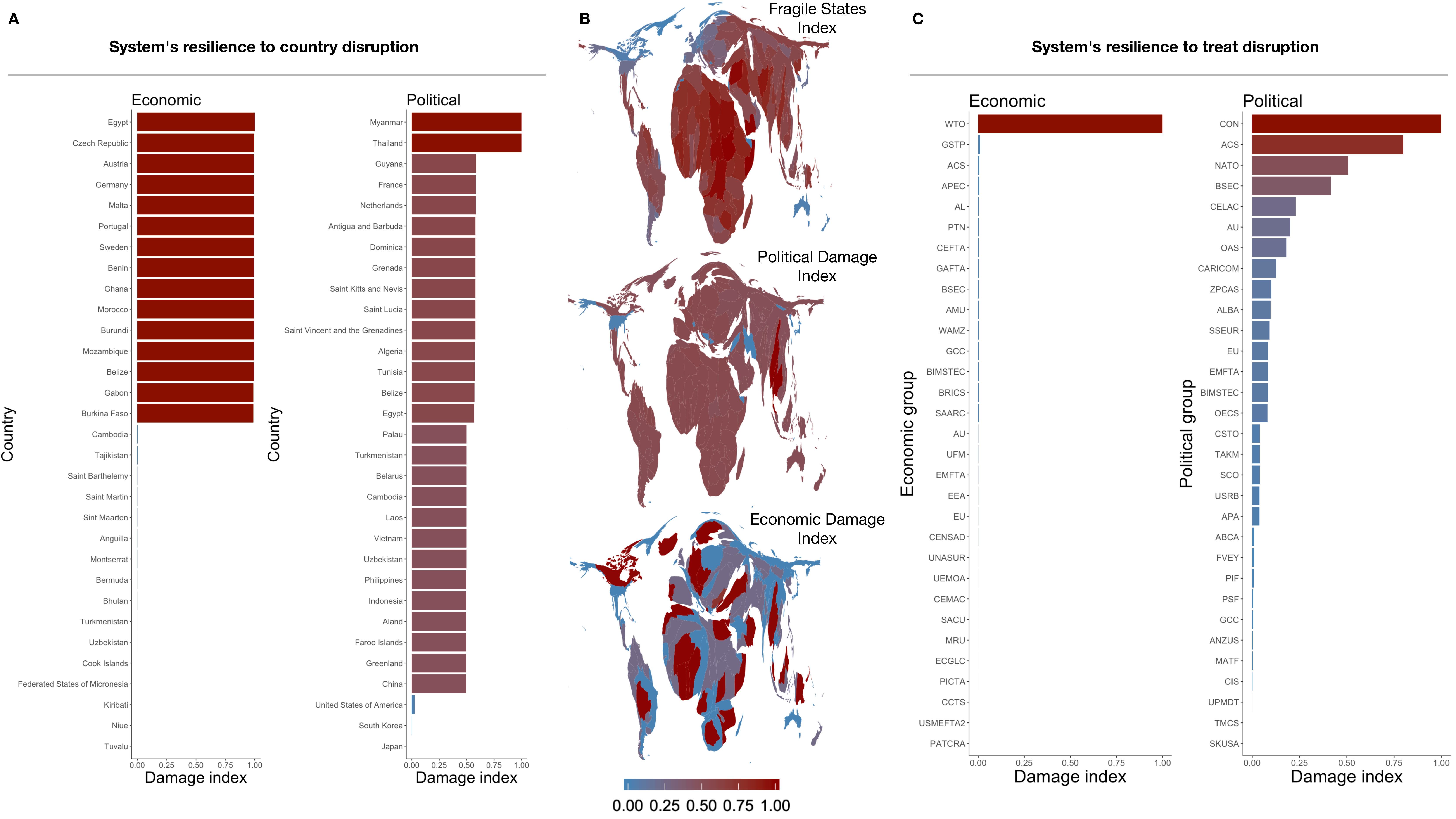}
\caption{\textbf{Cartography of world order resilience.} The central panel shows the distorted maps of Fragile States Index (FSI, top) vs Political Damage Index (PDI, middle) and Economic Damage Index (EDI, bottom), as defined in the text. Cartograms are built according to Gastner and Newman (33). In left and right panels we show the values of the damage index with respect to economic and political relationships, separately, in response to country-based disruptions (left) and treat-based disruptions (right): values range from 0 to 1, and we refer to the text for further details.}
\label{fig:cartogram}
\end{figure*}

Thus, the damage index is a graph structural descriptor of network disruption, and can be used to assess the resilience of the structural connectivity of the whole system. We have conducted several experiments on the world order network structure. We have computed the damage index in the economic layer (EDI), and the damage index in the political layer (PDI), for all possible individual disruptions, i.e. by removing a single country and its deals from one specific layer. We can therefore assess their impact on the whole world order structure. The damage index lies between 0 and 1, reaching the unit value for the most disruptive countries. The results of our analysis are depicted in Fig.~\ref{fig:cartogram}. We have made use of distorted maps \cite{gastner2004diffusion} to show the effect of every country's defection from economic or political deals upon the rest of the world. The distortion is computed for the damage index, and the color indicates its value for every country.

\paragraph{Independence of the damage index.} To disentangle the complementarity of our structural descriptor, the damage index, compared to other indices, we have selected the Fragile States Index (FSI), which is a measure of a State's internal sources of institutional, political, and socio-economic criticalities \cite{besley2011fragile}. A statistical analysis based on Spearman correlation reveals that there is no direct correlation between FSI and the two defined Economic (DIE;  $r = -0.10$, p-value~$= 0.18$) and Political Damage (DIP, $r = 0.05$, p-value~$= 0.53$) Indices, respectively. This result has important implications: it shows that intrinsically dysfunctional States need not be the most disruptive from a global perspective, as their \rev{turbulence} may be inherently local, whereas we consider the global impacts of a perturbation of the world order. It also implies that the world order is not critically susceptible to the factors that typically cause states to fail, e.g. localized corruption and low levels of human and socio-economic development.

\paragraph{Geo-political assessment.} The second important result is that the USA, as well as other major Western countries, are not disruptive, and the reason is that since Western countries tend to be tightly knit in their participation to international political or economic deals, the unilateral defection of a single country does not cause major structural damage as there is a significant level of redundancy. But the most surprising result comes from comparing PDI and EDI. In the political layer, the most critical countries are Myanmar and Thailand, even if there is a second group of countries that includes former colonial powers such as France and the Netherlands, and somewhat surprisingly a number of tiny countries which are in most cases overseas territories plus a few Northern African countries with important ties to Europe such as Algeria, Tunisia and Egypt, and a third group which includes among others China and more South-East Asian countries. It clearly looks paradoxical that a small island state such as, say, Saint Kitts and Nevis may have a (slightly) larger disruptive political potential than China. Moreover, the US damage index is extremely small. Overall, however, the size of the biggest disruption effects in the political layer is much smaller than in the economic one, although there is a large number of countries that are potentially critical. In the economic layer, instead, there are much less countries that are potentially critical, but the size of the associated disruptive effects is bigger than political ones. Moreover, the critical countries include a G7 member such as Germany, and also a number of smaller EU countries, of African countries, and Belize, which was also in the second group of politically disruptive countries; the same is true for Egypt, which is the only other country which is critical for both layers, and whose importance might be linked to its crucial mediation role in one of the most long-seeded, globally critical conflicts, the Israeli-Palestinian one \cite{schulz2019security}. Large disruption is therefore more likely to come from the global economy than from global politics. The politically most critical regions are South-East Asia, North Africa and Caribbean overseas territories, whereas the economically critical regions are Europe and Asia. Belize is a notable, common exception.

\section*{Discussion}

\paragraph{Country disruption potential.} Many African countries are found to have a major economically disruptive potential: Egypt, Benin, Ghana, Morocco, Burundi, Mozambique, Gabon, Burkina Faso. The only G7 country with high economic disruption potential is Germany, accompanied by (Czech Republic, Austria, Portugal, Sweden) and small (Malta) European countries. Finally, there is Belize. This is a surprising result, as most such countries are small or very small economies. The massive presence of African countries in this list seems to suggest that the economic world order still largely reflects the colonialist world order, a result that agrees with previous analyses carried out with different methodologies on different kinds of data \cite{buscema2015analyzing}, \cite{buscema2017kind}, \cite{erspamer2021global}. On the other hand, there is also a strong evidence that the German EU area of economic influence also plays a key role in the global architecture, as both Austria and the Czech Republic are economically disruptive. The presence of Portugal also hints at the ties with the colonial world order, given the important past of the country as a colonial power. Finally, we have small former colonies such as Malta and Belize.

Even more interesting, though, is the country disruption picture on the political side. Here, together with the already mentioned Myanmar and Thailand, we again find as already mentioned two past colonial powers such as France and the Netherlands. However, the most interesting feature of the list is the massive presence of micro-States, that is colonial overseas territories that are currently well recognized fiscal havens: Antigua, Dominica, Granada, Saint Lucia, Saint Kitts and Nevis, Saint Vincent. There are in addition three North-African countries (Egypt again, Algeria and Tunisia), Guyana and again Belize (two more overseas tax havens), plus a heterogeneous group of countries with lower but still substantial levels of disruptive capacity. Overall, the political picture draws a significant overlap of remnants of the colonial world order and global networks of tax avoidance. It is remarkable that the disruptive potential of fiscal havens is political, rather than economical. These results present interesting relationships with recent research that reconstructs the global architecture of tax evasion \cite{zucman2014taxing}, \cite{alstadsaeter2019tax}, suggesting that the political architecture of the world order could be still largely shaped by the primary capital accumulation of colonial empires \cite{sachs2020ages}, and the opaque system of tax havens which largely coincides with small territorial leftovers of previous, global territorial possessions \cite{haberly2015tax}, whose function was that of offshore sheltering of colonial riches during the uncertain decolonization process \cite{ogle2020funk}, and that have further consolidated their specialization ever since, also covering money laundering from illegal activities \cite{hampton2002offshore}.    

\paragraph{Treaties disruption potential.} On the other hand, the disruption picture from the point of view of international treaties is considerably simpler, to a surprising extent. The only treaty which has a major (and very substantial) disruptive capacity from the economic point of view is the free trade agreement administered by the WTO, whose multilateral governance role is tellingly undergoing significant changes \cite{hannah2018wto}. From the political point of view, the critical treaties are instead the Commonwealth of Nations, the Association of Caribbean States, NATO and Black Sea Economic Cooperation. The major role of the Commonwealth and of the Association of Caribbean States point attention again toward the colonial world order-tax evasion nexus. NATO has, for once, a rather intuitive role in the global architecture of political relations. The fact that the Black Sea Economic Cooperation has a major disruptive impact in political terms seems to stress the importance of the socio-economic stability of the Black Sea area which not incidentally has been at the center of major tensions between the E.U. and Russia in recent years.

\paragraph{Continuing legacy of the colonial world order.} The nonlinear structure of the global architecture of economic and political relations is surprisingly complex and yet at the same time informed by a familiar logic, and its critical conditions of robustness do not reflect the conventional wisdom. Despite its preliminary character, our analysis highlights how the legacy of the colonial world order, as reflected in the current vast network of overseas tax havens, seems to play a much bigger role in the shaping of the current world order than it could be imagined, to the point of turning the principle of country sovereignty of such micro-states into a pillar of a global system of tax evasion \cite{palan2020evolutionary}. The countries with the most disruptive potential are not, with few exceptions, major market democracies and institutional agents of democratic peace, but less developed countries (LDCs) whose role in the current world order is still largely determined by their colonial past, and overseas territories whose main economic specialization is the custody of large shares of the world's financial assets whose property is undisclosed \cite{haberly2015tax}. On the other hand, the disruptive potential of such countries is not merely related to corruption or lack of socio-economic development: as we have seen, neither political nor economic disruption is correlated to the FSI. It depends on the structural role they play in the architecture of the global governance system, which as we have seen it still largely shaped by a post-colonial logic. Interestingly, such logic also surfaces from the analysis of the disruptive potential of international treaties, with the additional driver of the East-West relations as a legacy of the Cold War and Russia-NATO antagonism. These unexpected structural features of the world order might explain, among other things, the persistence of difficulties in dismantling the global network of tax evasion despite the ambitious commitments of major world powers \cite{johannesen2014end}, \cite{konrad2016coordination}. Our results suggest that the structural weaknesses of the current world order could be due to causes that are different from the ones generally considered. Specifically, they seem to be much more fundamentally related than commonly thought to the opaque mechanisms that ensure perpetuation of global inequalities and post-colonial socio-economic divides.  This result will be of vital importance in re-thinking the world order after the COVID-19 crisis, as the pandemics has further, dramatically widened pre-existing income gaps both at national and global scales, so that a massive redistribution could be necessary to guarantee some minimal form of social justice  \cite{saez2021get}. So far, the incumbent global order has mostly adapted to the changing circumstances through tactical adjustment rather than major restructuring \cite{bisley2019contested}. But the policy agenda of the post-pandemic world might have to address fiscal injustice and its overlooked disruptive potential \cite{saez2019triumph} much more explicitly, while being called to move beyond the colonial world order and its still looming legacy for good.

\section*{Materials and Methods}

\begin{table}[!h]
\begin{footnotesize}
\caption{\label{tab:acronym} List of the geopolitical treaties considered in this study.}
\begin{tabular}{p{0.08\textwidth}p{0.35\textwidth}}
\textbf{Name} & \textbf{Description}\\\hline
AU            & African Union                                                                  \\ \hline
ACTO          & Amazon Cooperation Treaty Organization                                         \\ \hline
ABCA          & American-British-Canadian-Australian-New Zealand Armies                        \\ \hline
AC            & Andean Community                                                               \\ \hline
APA           & Anglo-Portuguese Alliance                                                      \\ \hline
AL            & Arab League                                                                    \\ \hline
AMU           & Arab Maghreb Union                                                             \\ \hline
AFTA          & ASEAN Free Trade Area                                                          \\ \hline
APEC          & Asia-Pacific Economic Cooperation                                              \\ \hline
APTA          & Asia-Pacific Trade Agreement                                                   \\ \hline
ACS           & Association of Caribbean States                                                \\ \hline
BRICS         & Association of emerging economies                                              \\ \hline
ASEAN         & Association of Southeast Asian Nations                                         \\ \hline
ANZCERTA      & Australia New Zealand Closer Economic Agreement                                \\ \hline
ANZUS         & Australia-New Zealand-United States Security Treaty                            \\ \hline
BA            & Baltic Assembly                                                                \\ \hline
BNS           & Baltic Naval Squadron                                                          \\ \hline
BIMSTEC       & Bay of Bengal Initiative for Multi-Sectoral Technical and Economic Cooperation \\ \hline
BENELUX       & Benelux union                                                                  \\ \hline
BSEC          & Black Sea Economic Cooperation                                                 \\ \hline
ALBA          & Bolivarian Alliance for the Peoples of Our America                             \\ \hline
CARICOM       & Caribbean Community                                                            \\ \hline
SICA          & Central American Integration System                                            \\ \hline
CEFTA         & Central European Free Trade Agreement                                          \\ \hline
CSTO          & Collective Security Treaty Organization                                        \\ \hline
CEZ           & Common Economic Zone                                                           \\ \hline
\end{tabular}
\end{footnotesize}
\end{table}

\begin{table}[!h]
\begin{footnotesize}
\begin{tabular}{p{0.08\textwidth}p{0.35\textwidth}}
\textbf{Name} & \textbf{Description}\\\hline
COMESA        & Common Market for Eastern and Southern Africa                                  \\ \hline

CIS           & Commonwealth of Independent States                                             \\ \hline
CISFTA        & Commonwealth of Independent States Free Trade Area                             \\ \hline
CON           & Commonwealth of Nations                                                        \\ \hline
CELAC         & Community of Latin American and Caribbean States                               \\ \hline
CENSAD        & Community of Sahel-Saharan States                                              \\ \hline
CCTS          & Cooperation Council of Turkic-Speaking States                                  \\ \hline
CAFTADR       & Dominican Republic-Central America FTA                                         \\ \hline
EAC           & East African Community                                                         \\ \hline
CEMAC         & Economic and Monetary Community of Central Africa                              \\ \hline
ECCAS         & Economic Community of Central African States                                   \\ \hline
ECGLC         & Economic Community of the Great Lakes Countries                                \\ \hline
ECOWAS        & Economic Community of West African States                                      \\ \hline
ECO           & Economic Cooperation Organisation                                              \\ \hline
EAEU          & EurAsian Economic Union                                                        \\ \hline
EMFTA         & Euro-Mediterranean Free Trade Area                                             \\ \hline
EEA           & European Economic Area                                                         \\ \hline
EFTA          & European Free Trade Association                                                \\ \hline
EU            & European Union                                                                 \\ \hline
FVEY          & Five Eyes/UKUSA                                                                \\ \hline
FPDA          & Five Power Defence Arrangements                                                \\ \hline
SSEUR         & Fourteen Eyes                                                                  \\ \hline
GSTP          & Global System of Trade Preferences among Developing Countries                  \\ \hline

GAFTA         & Greater Arab Free Trade Area / PAFTA                                           \\ \hline
GCC           & Gulf Cooperation Council                                                       \\ \hline
IOC           & Indian Ocean Commission                                                        \\ \hline
RIOPACT       & Inter-American Treaty of Reciprocal Assistance                                 \\ \hline
IAD           & Intergovernmental Authority on Development                                     \\ \hline
LHT           & Lancaster House Treaties                                                       \\ \hline
LAES          & Latin American Economic System                                                 \\ \hline
LAIA          & Latin American Integration Association                                         \\ \hline
MRU           & Mano River Union                                                               \\ \hline
MGC           & Mekong-Ganga Cooperation                                                       \\ \hline
MSG           & Melanesian Spearhead Group                                                     \\ \hline
EUROZONE      & Monetary union of EU                                                           \\ \hline
MATF          & Moroccan-American Treaty of Friendship                                         \\ \hline
\end{tabular}
\end{footnotesize}
\end{table}

\begin{table}[!h]
\begin{footnotesize}
\begin{tabular}{p{0.08\textwidth}p{0.35\textwidth}}
\textbf{Name} & \textbf{Description}\\\hline
NC            & Nordic Council                                                                 \\ \hline
NAFTA         & North American Free Trade Agreement                                            \\ \hline
NATO          & North Atlantic Treaty Organization                                             \\ \hline
OECS          & Organisation of Eastern Caribbean States                                       \\ \hline
GUAM          & Organization for Democracy and Economic Development                            \\ \hline
OAS           & Organization of American States                                                \\ \hline
TAKM          & Organization of the Eurasian Law Enforcement Agencies with Military Status     \\ \hline
OPEC          & Organization of the Petroleum Exporting Countries                              \\ \hline
PA            & Pacific Alliance                                                               \\ \hline
PICTA         & Pacific Island Countries Trade Agreement                                       \\ \hline
PIF           & Pacific Islands Forum                                                          \\ \hline
PATCRA        & Papua New Guinea-Australia Trade and Commercial Relations Agreement            \\ \hline
PSF           & Peninsula Shield Force                                                         \\ \hline
PTN           & Protocol on Trade Negotiations                                                 \\ \hline
SCO           & Shanghai Cooperation Organisation                                              \\ \hline
SAARC         & South Asian Association for Regional Cooperation / SAPTA                       \\ \hline
SAFTA         & South Asian Free Trade Area                                                    \\ \hline
ZPCAS         & South Atlantic Peace and Cooperation Zone                                      \\ \hline
SKUSA         & South Korea-United States Alliance                                             \\ \hline
SACU          & Southern African Customs Union                                                 \\ \hline
SADC          & Southern African Development Community                                         \\ \hline
MERCOSUR      & Southern Common Market                                                         \\ \hline
SICOFAA       & System of Cooperation Among the American Air Forces                            \\ \hline
TPSEP         & Trans-Pacific Strategic Economic Partnership                                   \\ \hline
TMCS          & Treaty of Mutual Cooperation and Security                                      \\ \hline

USMEFTA1      & U.S.-Middle East Free Trade Area                                               \\ \hline
USMEFTA2      & U.S.-Middle East Free Trade Area                                               \\ \hline
USMEFTA3      & U.S.-Middle East Free Trade Area                                               \\ \hline
USMEFTA4      & U.S.-Middle East Free Trade Area                                               \\ \hline
USMEFTA5      & U.S.-Middle East Free Trade Area                                               \\ \hline
UFM           & Union for the Mediterranean                                                    \\ \hline
UNASUR        & Union of South American Nations                                                \\ \hline
USRB          & Union State of Russia and Belarus                                              \\ \hline
UPMDT         & USA-Philippines Mutual Defense Treaty                                          \\ \hline
V4            & Visegrad Four                                                                  \\ \hline
UEMOA         & West African Economic and Monetary Union                                       \\ \hline
WAMZ          & West African Monetary Zone                                                     \\ \hline
WTO           & World Trade Organization                                                       \\ \hline
\end{tabular}
\end{footnotesize}
\end{table}



\paragraph{Data Availability.} \rev{The data that support the findings of this study are available from public repositories, without restrictions.}

\paragraph{Code availability.} \rev{The code used that support the findings of this study is available from the corresponding author upon reasonable request.}


\clearpage
\bibliographystyle{naturemag}
\bibliography{biblio}

\paragraph{Author contributions.} P.S. and M.D. performed the data analysis. All the authors designed the research and wrote the manuscript.

\paragraph{Competing interests.} The authors declare no competing interests.


\end{document}